# ROBUST MISSION DESIGN THROUGH EVIDENCE THEORY AND MULTI-AGENT COLLABORATIVE SEARCH

Massimiliano Vasile[1]
*Dipartimento di Ingegneria Aerospaziale, Politecnico di Milano*
*Via La Masa 34, 20156, Milan, Italy, vasile@aero.polimi.it*

**Abstract**

In this paper, the preliminary design of a space mission is approached introducing uncertainties on the design parameters and formulating the resulting reliable design problem as a multiobjective optimization problem. Uncertainties are modelled through evidence theory and the belief, or credibility, in the successful achievement of mission goals is maximised along with the reliability of constraint satisfaction. The multiobjective optimisation problem is solved through a novel algorithm based on the collaboration of a population of agents in search for the set of highly reliable solutions. Two typical problems in mission analysis are used to illustrate the proposed methodology.

## Introduction

In the early phase of the design of a space mission, the value of several design parameters is either unknown or is known with a degree of uncertainty. This uncertainty is either due to a lack of knowledge, and therefore is reducible in the further stages of the design, or is due to a stochastic variation of some quantities, and therefore is irreducible. The former kind of uncertainty is commonly said to be subjective or epistemic while the latter is defined as aleatory.

In recent times some researchers[1] have argued that it is inappropriate to represent all kinds of uncertainty solely by probabilistic means when enough information is not available. For this reason it has been proposed to use evidence theory (also referred to as Dempster-Shafer theory) for engineering applications and in particular for multidisciplinary design of aircraft.

In this paper it is proposed to apply evidence theory to the preliminary design of space missions in which uncertainty in subsystem sizing and orbit determination may effect the overall design. In particular performance indexes (such as useful payload mass, time of transfer, etc…) defining the optimal design point are affected by the uncertainty in the design parameters. It is therefore desirable to maximise reliability or belief in the optimum design point under this uncertainty. This can be done by translating a mission design problem into a multiobjective optimisation problem formulated in such a way to include uncertainty estimation effects both on constraint violation and on goal attainment. However the measures of uncertainty provided by evidence theory are discontinuous and nonsmooth functions and cannot be used with traditional gradient based optimizers because the sensitivities of the uncertain measures are not properly defined. In previous work this problem was solved resorting to smooth metamodels[2].

In this paper, instead, an approach based on a global multiagent collaborative search (MACS) hybridised with a deterministic domain decomposition techniques is used to find pareto optimal solutions that maximise reliability. This hybrid approach is an extension of evolutionary-branching (developed by the author in previous works[3,4]) to treat multiobjective problems. The evolutionary paradigm based on populations of chromosomes is here substituted with the idea of populations of agents with individual and social behaviours.

---

[1] Assistant Professor, Department of Aerospace Engineering, Politecnico di Milano

A particular decomposition technique is used to partition, or branch, the solution domain into subdomains containing the expected optimal value. By iterating alternatively branching and multiagent collaborative search the algorithms eventually converge reconstructing the Pareto front. Unlike other hybridisation of domain decomposition techniques and evolutionary algorithms, as the one proposed by Schutze et al.[6], here the partitioning mechanism is not purely systematic (bisecting each subdomain) but is adaptive, driven by the outcomes of the MACS and is not conceived to converge to the Pareto set but is used either to bound the set, pruning most of the search space, or to enhance diversity in the population of agents in order to ease their extensive exploration. This prevents the curse of dimensionality which would lead to an exponential growth of the number of subdomains with problem dimension. This hybridisation can be proven to be complete if an infinite number of subdivisions are allowed though in this case the algorithm has exponential complexity.

The proposed approach is applied to two test cases: aerocapture manoeuvres and low-thrust transfers.

## Uncertainty Modelling

It is a common habit, in reliable design, to classify uncertainties, whether it occurs in input parameters or in system model, into *aleatory* uncertainty and *epistemic* uncertainty. The former is also referred to as variability, irreducible uncertainty or stochastic uncertainty, while the latter is also referred to as reducible uncertainty or uncertainty due to a lack of knowledge. In order to correctly treat both kinds of uncertainty in a consistent computational framework for system engineering several authors have recently proposed the use of evidence theory[1,2]. Evidence theory, born with Dampster and Shafer in the '70, can be seen as a generalisation of probability and possibility theory. In the following we will use the notation presented in the work of Oberkampf and Helton[1].

Similar to probability theory, the evidence theory representation for the uncertainty in the design vector **x** is built up from the representation for the uncertainty in the components of **x**. Typically, in the early phase of the design process, specialists have a poor knowledge of some design parameters, therefore they express a *guess*. This guess is here represented by sets $\varepsilon$ of values a parameter can assume and a basic probability (BPA) $m(\varepsilon)$ is assigned to each set such that (i) $m(0)=0$ and (ii) $\sum_{\varepsilon \subset \Theta} m(\varepsilon) = 1$, where $\Theta$ is the set collecting all possible $\varepsilon$ and unions of $\varepsilon$. The BPA for **x** is computed with the simplifying assumption that there are no correlations or restrictions involving the components of **x**. Unlike probability theory that uses a single value as the only representation for uncertainty, evidence theory uses two values: the belief *Bel* and the plausibility *Pl* defined as follows:

$$Bel_y(Y_v) = Bel_D(f^{-1}(Y_v)) = \sum_{j \in I_B} m_D(\varepsilon_j) \tag{1}$$

$$Pl_y(Y_v) = Pl_D(f^{-1}(Y_v)) = \sum_{j \in I_P} m_D(\varepsilon_j) \tag{2}$$

where the two index sets $I_B$ and $I_P$ are defined as:

$$I_B = \{j : \varepsilon_j \subset f^{-1}(Y_v)\}; I_P = \{j : \varepsilon_j \cap f^{-1}(Y_v) \neq 0\} \tag{3}$$

Belief and Plausibility can be interpreted respectively as the minimum and maximum probability for $Y_v$ that is consistent with all the available evidence. The set $Y_v$ is defined as:

$$Y_v = \{y : y = f(\mathbf{x}) > v, \mathbf{x} \in D\} \tag{4}$$

and $\varepsilon_j \subset D$. Each possible value for $y$ is then given by all possible values assumed by the components of **x** within the subsets $\varepsilon_j$. The resulting BPAs for the product space $D = D_1 \times D_2 \times ... D_n$ of the subsets $\varepsilon_j$, is given by the Cartesian product $\varepsilon_1 \times \varepsilon_2 \times ... \varepsilon_n$ of the subset of the components of **x**:

$$m_C(\varepsilon_j) = \begin{cases} \prod_{i=1}^{n} m_{D_i}(\varepsilon_i) & \text{if } \varepsilon_i \subset D_i \text{ and } \varepsilon_j = \varepsilon_1 \times \varepsilon_2 \times ... \varepsilon_n \\ 0 & \text{otherwise} \end{cases} \tag{5}$$

This particular formulation requires to express a guess on the value of a design or model parameter without specifying any particular distribution of probability and without assigning any probability to any part of $\varepsilon$. Notice that a large interval with a high BPA represents a poor knowledge of the actual value of a parameter.

## Maximum Reliably Problem Formulation

An optimal design problem can be generally formulated as:

$$\min_{\mathbf{x} \in D} \mathbf{F}(\mathbf{x})$$

subject to: (6)

$$\mathbf{c}(\mathbf{x},\mathbf{p}) \leq 0$$

where $\mathbf{F}$ is vector of objective functions and $\mathbf{p}$ is a vector of parameters. The decision vector $\mathbf{x}$ contains the design parameters that have to be defined to optimise vector $\mathbf{F}$.

Once uncertainties are introduced, the optimal design problem has to be reformulated into a maximum reliability problem in order to take into account the degradation effects due to the uncertain parameters. Problem (6) can be re-casted in the following way:

$$\max_{\mathbf{x} \in D} Bel\left(\mathbf{F}(\mathbf{x},\mathbf{p}) \leq \mathbf{v}\right)$$

$$\min_{\mathbf{v} \in D} \mathbf{v}$$

$$\max \sigma \qquad (7)$$

subject to:

$$Bel(\mathbf{c}(\mathbf{x},\mathbf{p}) \leq 0) \geq \varepsilon$$

which states that the decision vector $\mathbf{x}$ has to maximise the belief in the optimal value $\mathbf{v}$ of the objective function $\mathbf{F}$ and in the satisfaction of the constraints $\mathbf{c}$, given the uncertainty margin $\sigma \in [0,1]$ and the uncertain parameter vector $\mathbf{p}$. The uncertainty margin is simply a scale factor applied to the width of the intervals containing the values of the uncertain parameters and is used to quantify how poor the knowledge of a particular quantity is.

## Optimisation Problem Formulation

Unlike single objective problems multiple objective problems look for a set of optimal values rather than a single optimal one. Anyway it has to be noticed that in many practical cases even single objective problems require the identification of multiple optimal solutions, in fact, if more than one solution exists within a required domain $D$ the interest could be more to find a number of feasible solutions forming a set $X$, rather than finding the global optimum with a high level of accuracy. Therefore the general problem, no matter if with a single or with multiple objectives, could be to find a set $X$ of feasible solutions $\mathbf{x}$ such that the property $P(\mathbf{x})$ is true for all $\mathbf{x} \in X \subseteq D$:

$$X = \{\mathbf{x} \in D \mid P(\mathbf{x})\} \qquad (8)$$

where the domain $D$ is a hyperparallelepiped defined by the upper and lower bounds on the components of the vector $\mathbf{x}$:

$$D = \{x_i \mid x_i \in [b_i^l, b_i^u] \subseteq \Re, i = 1,...,n\} \qquad (9)$$

All the solutions satisfying property $P$ are here defined to be optimal with respect to $P$ or $P$-optimal and $X$ can be said to be a $P$-optimal set. As an example in the case of multiobjective optimization, if $P$ is a dominance condition or Pareto optimality condition for the solution $\mathbf{x}$, then the solution is Pareto-optimal if $P(\mathbf{x})$ is true. In the case of single objective function, the set $X$ may contain all solutions that are local minimisers or are below a given threshold. Now property $P$ might not identify a unique set, therefore we can define a global optimal set $X_o$ such that all the elements of $X_o$ dominate the elements of any other $X$:

$$X_o = \{\mathbf{x}^* \in D \mid P(\mathbf{x}^*) \wedge \forall \mathbf{x} \in X \Rightarrow \mathbf{x}^* \prec \mathbf{x}\} \tag{10}$$

where $\prec$ represents the dominance of the $\mathbf{x}^*$ solution over the $\mathbf{x}$ solution. Now for multiobjective problems we can associate to each function vector $j$ a scalar dominance index $I_d$ such that:

$$I_d(\mathbf{x}_j) = \left| \{i \mid i \in N_p \wedge \mathbf{x}_i \succ \mathbf{x}_j\} \right| \tag{11}$$

where the symbol $|.|$ is used to denote the cardinality of a set and $N_p$ is the set of the indices of all the agents in the population. The property $P(\mathbf{x})$ in this case can simply define non-dominated solutions:

$$X = \{\mathbf{x} \in D \mid I_d(\mathbf{x}) = 0\} \tag{12}$$

## Multiagent Collaborative Search

Here a novel technique is proposed for the solution of problems (8) and (10). The idea is to associate each potential solution $\mathbf{x}$, represented by a string, of length $n$, containing in the first $n_I$ components integer values and in the remaining $n-n_I$ components real values, to an agent. An agent is an entity that can take actions, can communicate, has resources and can sense or perceive the environment.

Each agent is here endowed with a hypercube $\mathbf{S}$, enclosing a region of the solution space surrounding the agent and defined by a set of intervals $\mathbf{S}=S_1 x S_2 ... x S_n \subseteq D_l$, where $S_i$ contains the value of the component $x_i$.

The solution space is then explored locally by acquiring information about the landscape within each region $\mathbf{S}$ and globally using a population of agents.

A sequence of actions is performed by each agent according to a behavioural scheme $\beta_S$, in order to acquire a minimum set of samples sufficient to decide in which direction to take the next move. For an agent $\mathbf{x}_j$, a behavioural scheme is a collection of displacement vectors $\Delta \xi^s$ generated by a function $f_\beta$:

$$\beta_S = \{\Delta \xi^s \mid \forall s \in I_s \Rightarrow \mathbf{x}_j^{k+1} = \mathbf{x}_j^k + \Delta \xi^s \in D \wedge \Delta \xi^s = f_\beta(\mathbf{x}_j^k, \mathbf{x}_j^{k-1}, \mathbf{w}, \mathbf{r}, \mathbf{x}, s)\} \tag{13}$$

$$I_s = \{s \mid s \in N \wedge s \leq s_{\max}\} \tag{14}$$

where $f_\beta$ is a function of the current and past agent state $\mathbf{x}_j^k$ and $\mathbf{x}_j^{k-1}$, of a set of weights $\mathbf{w}$, a set of random numbers $\mathbf{r}$, the current state of the other agents in the population $\mathbf{x}$ and the index $s$. As an example, in particle swarm optimisation, the generating function $f_\beta$ can be defined as follows:

$$\Delta \xi^s = w_0(\mathbf{x}_j^k - \mathbf{x}_j^{k-1}) - w_1 r_1 (\mathbf{x}_j^k - \mathbf{x}_{j,lo}) - w_2 r_2 (\mathbf{x}_j^k - \mathbf{x}_{go}) \tag{15}$$

where the first and second components use local information of the agent while the third component introduces global information and is equivalent to a communication between the agent $\mathbf{x}_j$ and the best agent in the population $\mathbf{x}_{go}$. In the particular case in which the agent can perceive the environment locally, that is to say within $\mathbf{S}$, the behavioural scheme can be defined as follows:

$$\beta_S = \{\Delta \xi^s \mid s \in I_s \Rightarrow \mathbf{x}_j^k + \Delta \xi^s \in \mathbf{S}\} \tag{16}$$

$$I_s = \{s \mid s \in N \wedge s \leq s_{\max} \Rightarrow f(\mathbf{x}_j^k + \Delta \xi^s) \leq f(\mathbf{x}_j^k) + \varepsilon\} \tag{17}$$

where the set of indices $I_s$ is limited by a further condition on the value of the fitness function, i.e. the index $s$ is increased until a better solution is found and $s$ is not larger than a given $s_{max}$. $\beta_s$ defines the perception behavioural scheme. The behaviour scheme is conceptually equivalent to the pattern of common pattern search direct methods. In the present implementation the behavioural scheme is predefined by the user though a learning mechanism is under development for the adaptation of agent's behaviour depending on the problem under study.

In this work the behavioural scheme has an individualistic part and a cooperative part. For the individualistic part a random-walk strategy is used in conjunction with a linear and a quadratic local model of the solution space. The random-walk strategy generates a displacement randomly perturbing the location of the agent with a nonuniform distribution (this is equivalent to a nonuniform mutation). The new sampled point is then used to construct a linear unidimensional model. With this model a new displacement is generated and a new sample is memorised. This second sample is then used to build a quadratic model of the function. This scheme is repeated $s$ times and then the direction of maximum improvement registered in the past history of the agent is included in the list as an additional displacement.

This behaviour is rather individualistic since does not exploit any information coming from other agents. In order to exploit this information the additional displacement is sometime computed taking a direction suggested by another agent. All the collected samples are then compared and only the subset showing some improvement is retained.

The contraction or expansion of each region $S$ is regulated through a scaling factor $\rho$, applied to the dimensions of $S$, which depends on the findings of the perception mechanism: if none of the samples is improving the agent's status, the radius is contracted up to the distance from the agent of the best sampled point. When $\rho<\rho_{min}$ the agent is said to be converged and is regenerated. Moreover if many agents are intersecting their migration regions and their reciprocal distance falls down below a given threshold, the worst one is regenerated.

Since local perception is a costly operation, sampling behavior is limited by the level of resources allocated to each agent. The level of resources is $s_{max}$ and is simply decreased of one unity if no improvement is registered from generation $k$ to generation $k+1$ and increased of one unity if an improvement is registered. The minimum level of resources has been fixed to 1 and the maximum level to problem dimensions.

Moreover the perception behavior is applied to a portion $P_f$ of the entire population. The size $n_f$ of this subpopulation is fixed and a filter operator is used to fill in the set $P_f$. The filter ranks all the agents on the basis of their fitness from the best to the worst and then selects the first $n_f$, the remaining agents are either hibernated (i.e. no operator is applied) or mutated. The probability of being mutated or hibernated depends on their ranking.

**Collaboration, Knowledge Sharing and the Global Archive**

At every stage of the optimisation process, a number of solutions (i.e. of agents) belongs to an optimal set $X$. This set is saved into a global archive $A_g$ which represents also the repository of the best achieved and perceived locations of each agent and of the location and fitness of the converged agents.

After the behavioural scheme is applied to all the agents in $P_f$, the ones that present an improvement are inserted in a communication list and communication is performed between each element of the list and an equal number of agents randomly selected form the population.

Moreover the elements in $A_g$ are ranked according to their crowding factor (the most isolated element is the least crowded) and added to the communication list. The crowding factor is computed in the following way: each agent swipes through the archive $A_g$ and, if not dominated, adds to each element in $A_g$ the value of the inverse of the distance from it, divided by the total number of elements in the archive. After the archive has been ranked the elements in $A_g$, from the most isolated to the most crowded, communicate a moving direction, a displacement $\Delta\xi$, to all the agents that are either dominated or have shown no improvements from the previous generation.

**Constraint Handling Technique**

The algorithm described solves bound-constrained problems but since in most of the cases, constraints are nonlinear, an extension of the algorithm has been developed in order to take into account nonlinear inequality constraints. At each generation the population of agents is divided into two subpopulations and a different objective function is assigned to each one, namely one subpopulation aims at minimising the original objective function while the other aims at minimising the residual on the constraints defined as:

$$\min_{y \in D} f = \sum_{j=1}^{m} \max([0, R_j]) \tag{18}$$

The two subpopulations are evolved in parallel and agents are allowed to jump from one population to the other, i.e. if a feasible agent becomes infeasible it is inserted in the subpopulation of infeasible agents and assigned to the solution of the constraints, on the other hand if an infeasible agent becomes feasible it is inserted in the population of feasible individuals and allocated to the minimisation of the original bound constrained objective function $f$. As a result the final optimal solution either is feasible or minimises infeasibilities. This procedure does not maintain feasibility for any agent, therefore once a feasible set has been found the perception mechanism is used to ensure that every move maintains the feasible population inside the feasible set. If $f^*$ is the value of the objective function of an agent **x** inside the feasible set, the objective function of a new agent generated from **x** is then augmented with the maximum among the residuals **R** on the constraints:

$$\min_{y \in D} f = \begin{cases} f^* & \text{if every } R_j \leq 0 \\ f^* + \max \mathbf{R} & \text{if any } R_j > 0 \end{cases} \tag{19}$$

The described strategy co-evolving two populations with two different goals, allows a flexible search for feasible optimal solutions: in fact through the described use of the perception mechanism feasibility can be enforced on all feasible solutions. In this case the exploration of the solution space may be over penalised reducing the convergence rate. Therefore in order to search more extensive along the boundary of the feasible region a subset of the feasible solutions is allowed to temporary violate the constraint while preserving the feasibility of at least the best solution.

## Domain Decomposition

In order to improve search space exploration, the initial domain $D$ is progressively decomposed into smaller domains $D_l \subseteq D$ according to a branching scheme. Before each run of MACS the solution space is partitioned according to a prediction. During MACS the number of sampled points within each subdomains is computed and assigned as a density value to the subdomain.
When an agent is either regenerated or mutated its new location is taken within the subdomain with the lowest density value in order to favor and homogenous spreading of the samples .
Branching is based on the output of the MACS step and produces a decomposition of the solution space $D$ into subdomains $D_l$ such that:

$$\bigcup_{l=1}^{M} D_l = D \tag{20}$$

Decomposition (20) is then iteratively applied to the subdomains $D_l$ that need further exploration so that:

$$\bigcup_{l=1}^{M} D_l^{(d)} = D_l^{(d-1)} \tag{21}$$

where $d$ is the branching or decomposition depth.

### Branching Scheme

A branching scheme is represented by a set $I_s$ containing the indices of the coordinates that have to be split and a set $C_B$ containing the cutting point for each coordinate. The initial set $I_s$ is defined by the user while the set $C_B$ contains the middle point of the interval defining each coordinate. After each run of MACS the branching scheme is adapted depending on the outcome of the MACS. If a coordinate has a high number of clusters and a low number of cuts then it is included in the set $I_s$. The cutting point is then recomputed as the middle point between the cluster containing the best fitness and the cluster containing the worst fitness. If just one cluster exists then one of the boundaries is used instead of the other cluster.

The subdomains $D_l$ with the least number of collected samples among the ones containing elements of $X$ is selected for further decomposition provided that the number of times $n_b$ its parent subdomains have been branched without improvement is below a given threshold. This strategy however excludes from further exploration all the subdomains containing no elements of $X$. Therefore when an exhaustive search is required, the following merit function is used:

$$\psi_{D_l} = (1-\upsilon)\varpi_{D_l} + \upsilon\varphi_{D_l} \qquad (22)$$

where $\varpi_{D_l}$ is the density of function evaluations in $D_l$, $\varphi_{D_l}$ is best fitness in $D_l$ and $\upsilon$ is a weighting factor used to favor either convergence or exploration. For multiobjective problems, in the following, we use the former strategy.

## Standard Multiobjective Test Cases

The proposed optimization approach, implemented in a software code called EPIC, is here tested on two standard problems, well known in the literature, the details about these test functions are summarized in Tab.1. Test case *DEB* is a constrained multiobjective optimization problem, proposed by Deb[11], with a convex Pareto front. The constraint $C$ fragments the admissible Pareto front into a disconnected set. Running Epic with a single level of branching and a limited population of 10 agents with 5 explorers, iterated for 12000 function evaluations, led to the result reported in Fig.1 where the constraint is represented with a continuous line while the solution found by EPIC is reported in dotted line.

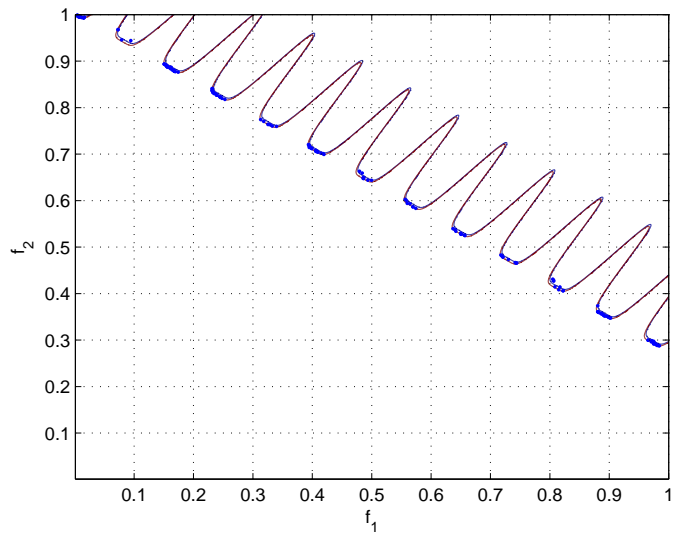

Figure 1. Admissible Pareto front for problem *DEB*.

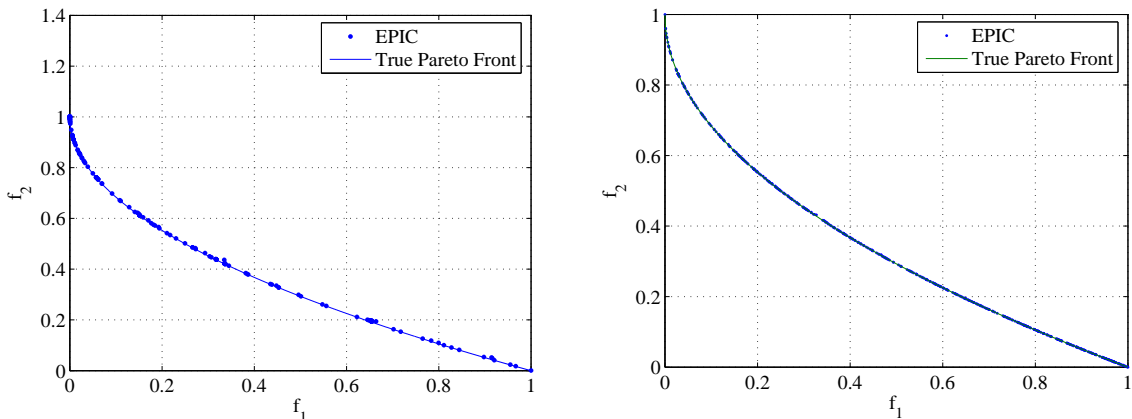

Figure 2. Optimal Pareto front for problem ZDT4, 3 exploring agents running for 10000 function evaluations, on the left, and running for 20000 function evaluations, on the right

Table 1. Multiobjective test functions

| ID | PARAMETERS | OBJECTIVE FUNCTION AND CONSTRAINT |
|---|---|---|
| DEB | $n=10$<br>$x_i \in [0,1]$<br>$i=1,...,n$<br>$a=0.2; b=10$<br>$e=1; c=1$<br>$d=6$ | $f_1 = x_1$<br>$f_2 = g\left[1 - \sqrt{\frac{f_1}{g}}\right] + 1; \quad g = 1 + \frac{9}{n-1}\sum_{i=2}^{n} x_i$<br>$C = \cos\theta(f_2 - e) - f_1 \sin\theta - a\left|\sin(b\pi(\sin\theta(f_2-e) + f_1\cos\theta)^c)\right|^d$ |
| ZDT4 | $n=10$<br>$x_1 \in [0\ 1]$<br>$x_i \in [-5\ 5]$<br>$i=2,...,n$ | $f_1 = x_1$<br>$f_2 = g\left[1 - \sqrt{\frac{f_1}{g}}\right]$<br>$g = 1 + 10(n-1) + \sum_{2}^{n} x_i^2 - 10\cos(4\pi x_i)$ |

Test *ZDT4* is commonly recognized as one of the most challenging problems since it has $21^9$ different local Pareto fronts of which only one corresponds to the global Pareto-optimal front. In this case the exploration capabilities of each single agents are enough to locate the correct front with a very limited effort. In fact even with just five agents it was possible to reconstruct (see right plot in Fig. 2) the correct Pareto front 20 times over 20 different runs. The total number of function evaluations was fixed to 20000 for each of the runs, though already after 10000 function evaluations EPIC was always able to locate the global front (see left plot in Fig. 2). Despite the small number of agents the sampled points of the Pareto are quite well distributed with just few and limited interruptions. The use of a limited number of agents instead of a large population is related also to the complexity of the algorithm. In fact the complexity of the procedure for the management of the global archive is of order $n_A(n_p+n_A)$, where $n_A$ is the archive size and $n_p$ is the population size, while the complexity of the exploration-perception mechanism is of order $n_p(n+n_p)$, therefore, even if the algorithm is overall polynomial in population dimension, the computational cost would increase quadratically with the number of agents. As an additional proof of the effectiveness of MACS we compare the average euclidean distance of 500 uniformly spaced points on the true optimal pareto front from an equal number of points belonging to the solution found by EPIC with the analogous performance metric found in [9] for NSGA-II, SPEA and PAES (see Table 2).

Table 2. Mean and standard deviation of the distance metric

| APPROACH | AVERAGE EUCLIDEAN DISTANCE | STANDARD DEVIATION |
|---|---|---|
| MACS | 1.542e-3 | 5.19e-4 |
| NSGA-II | 0.513053 | 0.118460 |
| SPEA | 7.340299 | 6.572516 |
| PAES | 0.854816 | 0.527238 |

## Robust Low-thrust Transfer

The first space related test case is a very simple problem of low-thrust transfer design aiming at the minimization of the propellant mass required to reach a given target. Since the typical parameters characterizing the propulsion system are generally poorly known in the preliminary phase of the design it is customary to parameterize the optimization for several values of $I_{sp}$, thrust level and specific power. The last parameter has a considerable importance on the overall mass of the spacecraft since the mass of the power system could make the gain in propellant irrelevant.

The trajectory and the control profile are here computed with a shape-based approach representing the evolution of the nonsingular equinoctial orbital parameters *[p,f,g,h,k]*, with respect to the true longitude *L*, through an exponential shape[5]:

$$\boldsymbol{\alpha} = \boldsymbol{\alpha}_0 + \boldsymbol{\alpha}_1 e^{\boldsymbol{\alpha}_2(L-L_0)} \quad (23)$$

where the two boundary conditions vectors $\boldsymbol{\alpha}_0$ and $\boldsymbol{\alpha}_1$ depends on the boundary conditions on position and velocity while the shaping vector $\boldsymbol{\alpha}_2=[\alpha_{21}, \alpha_{22}, \alpha_{23}]$ can be used to optimize the cost function associated to the trajectory. The required control acceleration profile can be computed from:

$$\mathbf{a}_d = \ddot{\mathbf{r}}(\boldsymbol{\alpha}) + \mu \frac{\mathbf{r}(\boldsymbol{\alpha})}{r(\boldsymbol{\alpha})^3} \quad (24)$$

Notice that only three shaping parameters are introduced, though the equinoctial elements are five, one for the semi-latus recto *p*, one for *f* and *g* and one for *h* and *k*. The main reason for this choice is that the shaping parameters change the curvature and then the rate of growth of the orbital elements. Now both *f* and *g* depends linearly on the eccentricity and are periodical function of the anomaly of the pericentre and of the argument of the ascending node while *h* and *k* are periodical function of the argument of the ascending node and both depend on the tangent of the inclination. Therefore for these four elements the rate of growth on the long term depends only on two parameters, namely the eccentricity and the inclination.

After the control profile is available, the mass consumption is computed with:

$$m_p = 1 - e^{-\int_{L_0}^{L_f} \frac{|\mathbf{a}_d|}{I_{sp} g_o} \frac{dt}{dL} dL} \quad (25)$$

with the additional constraint on the time of flight:

$$T = \int_{L_0}^{L_f} \frac{dt}{dL} dL \quad (26)$$

The maximum deliverable thrust is a simple function of the inverse of the square of the distance from the Sun, of the surface area of the solar array *A*, of the specific power *w* and of the overall efficiency of the power system and of the engine $\eta_p$:

$$\Phi_{\max} = \eta_p \frac{wAp_0}{r^2} \quad (27)$$

with the specific impulse estimated to be a linear function of the specific power:

$$I_{sp} = \frac{2\eta_e w}{g_0} \quad (28)$$

The total mass of the spacecraft *m* and the mass of the power system $m_{SA}$ are simply:

$$\begin{aligned} m &= m_p + m_{SA} + m_s; \\ m_{SA} &= 1.1 A \rho_{SA} \end{aligned} \quad (29)$$

where $m_s$ is the total mass of the remaining parts of the spacecraft. Now if we consider an uncertainty on power efficiency conversion $\eta_p$, on the power per unit area at 1AU $p_0$, and on the engine efficiency $\eta_e$, the problem can be reformulated as follows:

$$\min \begin{cases} 1 - Bel(m_p + m_{SA} < m_{\max}) \\ -m_{\max} \end{cases}$$

*subject to* (30)

$$ma_d \leq \Phi_{\max}$$

$$t_f - T \leq \varepsilon_T$$

The control parameters for the multiobjective search are then:

$$\mathbf{x} = [t_0, t_f, w, A, n, \alpha_{21}, \alpha_{22}, \alpha_{23}, m_{\max}] \tag{31}$$

The range of the uncertain parameters and their associated elementary probability assignments are summarized in Tab.3, while the search domain for the design parameters is represented in Tab. 4. The range of uncertainty of the specific power is related to the uncertainty on the efficiency of the solar cells and is reducible. In this case a guess was expressed on the probability of having Si cells or GaAs cells. On the other hand the uncertainty on efficiency of the overall power system is more difficult to predict due to the several losses of cabling and harness, thus a guess was expressed in order to comprise the minimum and maximum values known empirically.

*Table 3. Ranges of uncertain quantities*

| Uncertain Parameter | $BPA_1$ | $BPA_2$ |
|---|---|---|
| $\eta_p$ | 1.0 [0.77 0.98] | - |
| $p_0\ (W/m^2)$ | 0.65 [251.23 300.12] | 0.35 [258.02 349.01] |
| $\eta_e$ | 0.7 [0.6 0.65] | 0.3 [0.65 0.75] |

*Table 4. Solution domain for the low-thrust problem*

|    | $t_0$(MJD2000) | $t_f$(day) | w | $A(m^2)$ | N | $\alpha_{21}$ | $\alpha_{22}$ | $\alpha_{23}$ | $m_{\max}$ |
|---|---|---|---|---|---|---|---|---|---|
| Ub | 7000 | 1000 | 35 | 25 | 1 | -1 | -1 | -1 | 1 |
| Lb | 3650 | 640 | 16 | 1 | 2 | 1 | 1 | 1 | 0 |

The maximum belief Pareto front for problem (30) has been plotted in Fig. 3, notice how the front is made of two classes of solutions corresponding to short and long transfer times. In both cases Fig. 4 demonstrates how the solutions are robust against the whole range of possible variations of the thrust. Fig. 5 shows an interesting property of the proposed shape-based approach: non singular equinoctial elements have been propagated forward in time with the control low found by the proposed shape-based method and compared to the corresponding shaped elements. As can be seen the shaped elements behave as averaged elements.

Finally Fig.6 represents two sample trajectories for the long and short transfer options: arrows represent the thrust vector along the trajectory.

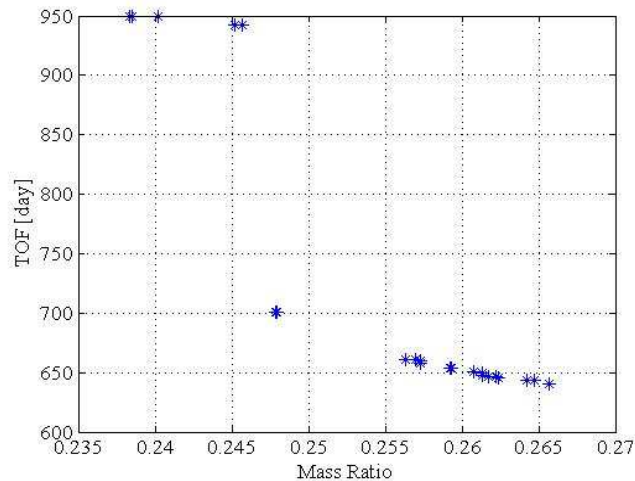

Figure 3. Pareto set of robust solution with maximum Belief

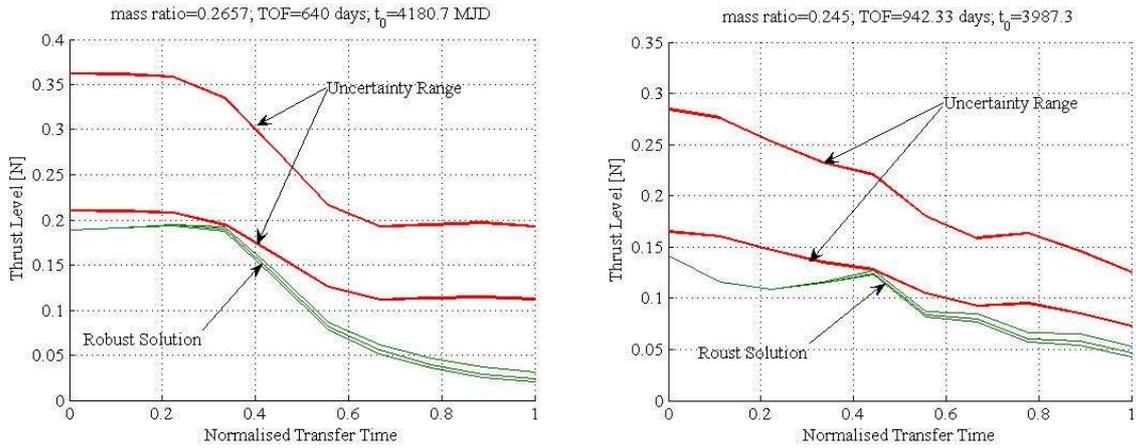

Figure 4. Robust thrust profile for short, on the left, and long, on the right, transfer options

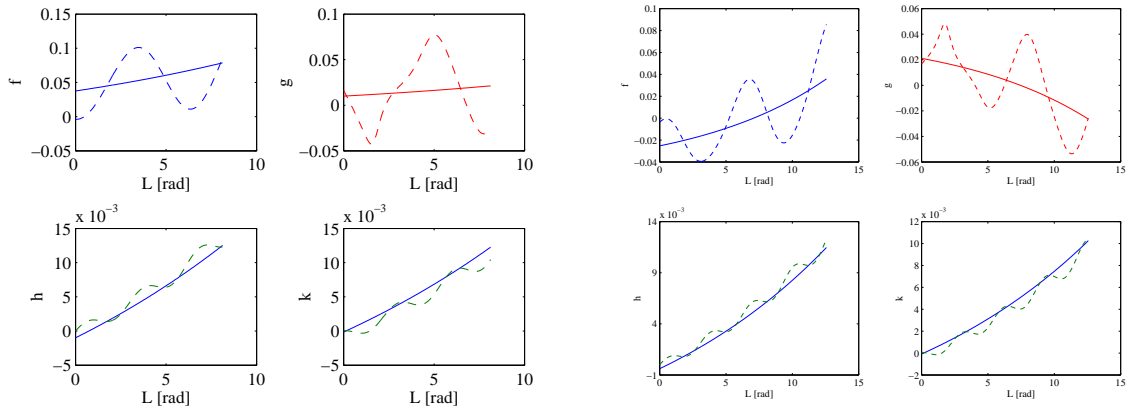

Figure 5. Propagated (dashed line) and shaped-equinoctial elements (solid line) for short, on the left, and long, on the right, transfer options

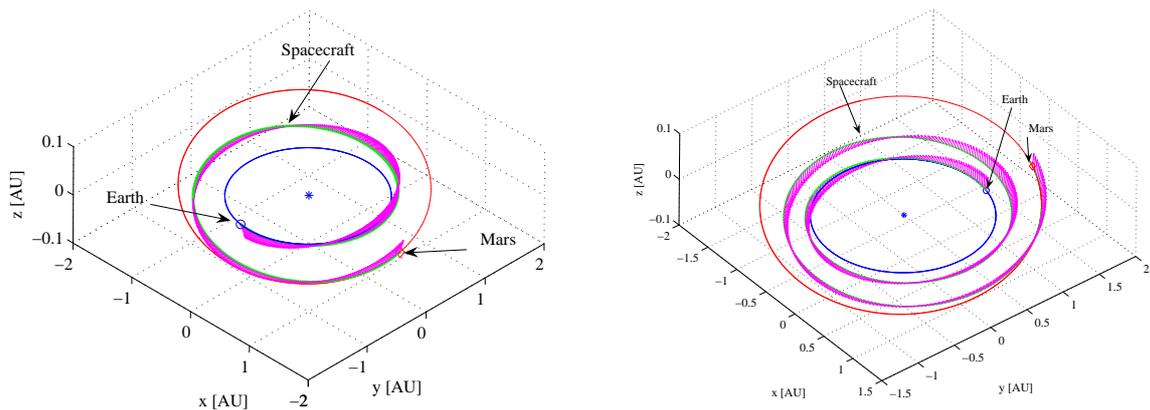

Figure 6. Short, on the left, and long, on the right, transfer options from Earth to Mars

## Robust Aerocapture Design

Aerocapture manoeuvres are generally composed of two phases: an atmospheric phase and an orbital phase. The orbital phase before the atmospheric phase is propagated analytically from the arrival point on the sphere of influence to the entry point in the atmosphere. The atmospheric phase is then propagated numerically using the following set of differential equations[8]:

$$\dot{r} = v \sin \beta$$
$$\dot{\theta} = \frac{v \cos \beta \sin \chi}{r \cos \psi}$$
$$\dot{\psi} = \frac{v \cos \beta \cos \chi}{r}$$
$$\dot{v} = -g \sin \beta - \frac{D}{m}$$
$$\dot{\chi} = \frac{v \cos \beta \sin \chi}{r} \tan \psi + \frac{L \sin \nu}{v \cos \beta}$$
$$\dot{\beta} = -\frac{g \cos \beta}{v} + \frac{L \cos \nu}{v} + \frac{v \cos \beta}{r}$$

(32)

Aerodynamic forces $D$ and $L$ are modelled as follows:

$$D = \frac{1}{2} \rho(H,h) S C_d(C_l) v^2$$
$$L = \frac{1}{2} \rho(H,h) S C_l v^2$$

(33)

where $C_l$ is the lift coefficient and is considered constant during the entire atmospheric phase, $C_d$ is the drag coefficient, $S$ the reference surface of the spacecraft and $m$ its mass. Here the atmospheric density $\rho$ follows an exponential profile function of the altitude $h$ parametrised in the density scale height $H$. A simple drag polar function is used for the drag and lift coefficients[7]:

$$C_D = C_{D_0} + \frac{C_{D_0}}{(n-1)C_{l_{max}}^n} C_L^n$$

(34)

where the maximum lift coefficient $C_{lmax}$ and the exponent $n$ depend on the shape and on the Mach number. Moreover, for a blunt body(see Fig.7), the zero-lift drag coefficient $C_{D0}$ can be expressed as a function of the geometric properties of the spacecraft, namely the ratio between the nose radius of curvature $R_n$ and the radius of the cross section area $R_b$ and the half cone angle $\Theta$[7][8]:

$$C_{D_0} = C_{pt2} \left[ \sin^2 \Theta \left( 1 - \left(\frac{R_n}{R_b}\right)^2 \cos^2 \Theta \right) + \frac{1}{2}\left(\frac{R_n}{R_b}\right)^2 \left(1 - \sin^4 \Theta\right) \right]$$

(35)

for hypersonic flow, and following the modified Newtonian theory, the total pressure coefficient can be expressed as a function of the ratio between the specific heats of the gas $\gamma$ and the Mach number $M_\infty$:

$$C_{p,t_2} = \frac{2}{\gamma}\left(\frac{\gamma+1}{2}\right)^{\frac{\gamma}{\gamma-1}} \left[ \frac{\gamma+1}{\left(2\gamma - \frac{\gamma-1}{M_\infty^2}\right)} \right]^{\frac{1}{\gamma-1}} - \frac{2}{\gamma M_\infty^2}$$

(36)

The Mach number $M_\infty$ is a function of the gas properties, however for the hypersonic flight regimes typical of aerocapture manoeuvres, the pressure coefficient is almost constant with the Mach number. Therefore the temperature $T_\infty$ and the gas constant $R$ will not be included among the uncertain parameters. Moreover, for a blunt body with rotational symmetry, the critical lift, for which the lift-to-drag ratio is maximum, can be expressed as a function of the $C_{D0}$ and of the exponent $n$:

$$C_{L_{max}} = \frac{(n-b)n C_{D_0}}{a(n-1)}$$

(37)

where *a* and *b* are generally taken from an empirical fit[7]. Post atmospheric conditions are then corrected with three manoeuvres, two at the apocentre $\Delta v_2$ and $\Delta v_i$ to adjust the pericentre and change the inclination and two at the pericentre $\Delta v_1$ and $\Delta v_3$ of the orbit resulting from the aeromanoeuvre in order to adjust the apocentre. The first manoeuvre at the pericentre is essential to correct post aeromanouevre conditions with a hyperbolic velocity. The total mass consumption due to the corrective manoeuvre can be computed with the simple rocket equation:

$$m_p = m\left(1 - e^{-\frac{\Delta v_1 + \Delta v_2 + \Delta v_3 + \Delta v_i}{I_{sp} g_0}}\right) \qquad (38)$$

where the specific impulse has been taken equal to $I_{sp}=350s$. According to this formulation of the cost function a robust solution would manifest a low mass consumption and minor corrective manoeuvres.

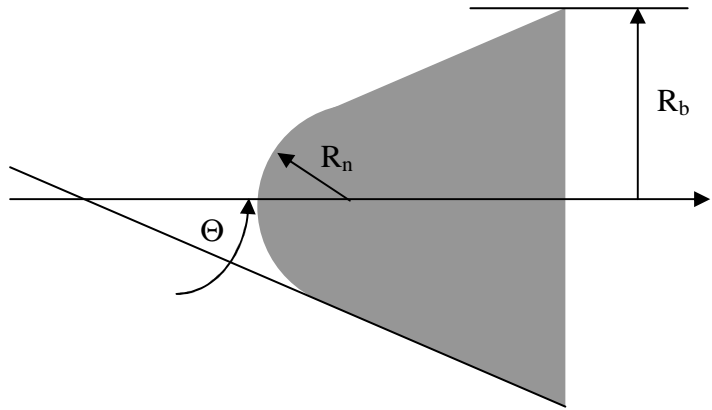

Fig. 7. Cartoon of the Spacecraft model

Due to the high sensitivity of aerocapture manoeuvres, an optimal design of the associated trajectory would require a high accuracy in the initial conditions and spacecraft aerothermodynamic properties. It is a common use to analyse the problem looking at the width of the entry corridor, i.e. the interval of entry angles for which the exit orbital parameters are as desired and the maximum g-load and heat flux are under a given threshold. This generally requires some control capabilities of the spacecraft during the atmospheric phase. In this paper we tackle the problem from a different prospective. In case the control is limited or unknown, the spacecraft design should be as such to allow aerocapture even for uncertain parameters and initial conditions given a fixed value of the control variables. The problem can be generally formulated as follows:

$$\begin{aligned}\min_{\mathbf{x}\in D} \; & m_p \\ & q(\mathbf{x}) \leq q_{\max} \\ & n_g(\mathbf{x}) \leq n_{\max}\end{aligned} \qquad (39)$$

All feasible solutions must guarantee that the heat flux $q(\mathbf{x})$ and the load factors $n_g(\mathbf{x})$ remain below $q_{max}$ and $n_{max}$ respectively, for the entire atmospheric phase. The heat flux along the trajectory is computed with the equation:

$$q = k_h \sqrt{\frac{\rho}{R_n}} v^3 \qquad (40)$$

where $k_h$ has the typical[10] value 1.89e-8 if the heat flux is expressed in W per cm$^2$.

Here uncertainties in the value of atmospheric density and gas thermal properties are considered epistemic due to a lack of information. On the other hand uncertainties on the state vector at the entrance of the atmosphere can be regarded either as aleatory, if an accurate orbit determination is performed, or as epistemic, during the preliminary design phase. Other sources of uncertainties related to spacecraft system parameters, such as shape, geometry and aerodynamic coefficients are considered to be epistemic.

Table 5 summarises all the uncertain parameters with their associated BPA and interval of uncertainty. It is here assumed that the BPA have been derived from a data fusion process collecting different opinions from several experts and that has assigned a certain level of confidence to certain ranges, moreover it is assumed that the interval of values for $n$ corresponds to experimental data for all possible shapes and configurations that can be generated with model (35). Notice that the uncertainty on the velocity and on the flight path angle are relative quantities since the two parameters are decisional variables moreover their BPA for this test case, has been set to 0.99 assuming that the complement to infinity would contain the remaining 0.01 of probability.

Notice that the variability of $H$ is quite consistent ranging from 5 to 12 and the basic probability assigned to the interval [7 12] is almost equal to the probability assigned to the interval [8 10]. This poor knowledge of $H$ is reducible but in the early phase of the design will give a measure of the sensitivity to this parameter.

*Table 5. Ranges of uncertain quantities*

| Uncertain Parameter | $BPA_1$ | $BPA_2$ | $BPA_3$ |
|---|---|---|---|
| $H$ (km) | 0.1 [5 7] | 0.4 [7 12] | 0.5 [8 10] |
| $\rho_0$ (kg/cm$^3$) | 0.2 [0.15 0.2] | 0.8 [0.19 0.21] | - |
| $\gamma$ | 0.1 [1.2 1.25] | 0.9 [1.25 1.3] | - |
| $n$ | 1 [2.0 2.15] | - | - |
| $\delta v$ (km/s) | 0.99 [-1 1] | - | - |
| $\delta \beta$ (deg) | 0.99 [-1 1] | - | - |
| $\delta \chi$ (deg) | 0.99 [-1 1] | - | - |

Now problem (39) can be reformulated as a reliability optimisation problem: the aim is to find the optimal design parameters $S$, $\vartheta$, $R_n/R_b$ and the control parameter $v, \beta$, $\upsilon$ and $C_l$ in order to maximise the belief of being captured with a $\Delta v$ correction below a given value. Here the propellant mass is used as a performance index of the aeromanoeuvre. Problem (39) then becomes:

$$\mathbf{F}(\mathbf{x}) = \begin{cases} 1 - Bel_r(m_p < m_{\max}) \\ -(\sigma_1 + \sigma_2) \\ m_{\max} \end{cases} \quad (41)$$

subject to:

$$Bel(q \leq q_{\max}) > 0.99; \\ Bel(n_g \leq n_{\max}) > 0.99; \quad (42)$$

where constraint equations (36) are substituted with a condition on the belief that constraints are always satisfied for any aerocapture manoeuvre. Here the $q_{max}$ is taken equal to 50 W/cm$^2$ and the maximum load factor $n_{max}$ is fixed at 5g. The design state vector is made up of the design parameters, the control variables plus what could be considered the design margin $\sigma_1$ and $\sigma_2$ on the estimated error on the entry angle and entry velocity respectively :

$$\mathbf{x} = [v, \beta, \nu, C_l, \vartheta, S, R_n/R_b, m_{\max}, \sigma_1, \sigma_2]^T \quad (43)$$

The parameter $R_b$ is the ratio the radius of the reference frontal surface between and $R_n$ the nose tip radius. The solution domain defined by the ranges of each component of the design vector are summarized in Tab.6. This solution space constrains the optimisation to blunt bodies with a maximum half angle of 40 degrees and a minim arrival velocity above the parabolic limit. A second run was then performed with an extended set of values in order to include both parabolic (the lower limit) and hyperbolic arrival conditions at Mars. In this case the half cone angle of the blunt body is free to vary from 5 to 60 degrees and $R_n/R_b$ from 0.1 to 0.9. This second solution space is reported in Tab. 7.

*Table 6. Solution domain for the aerocapture problem*

|    | $v(km/s)$ | $\beta\,(deg)$ | $\upsilon$ | $C_l$ | $\vartheta\,(deg)$ | $S\,(m^2)$ | $R_n/R_b$ | $m_{max}$ | $\sigma_1$ | $\sigma_2$ |
|----|-----------|----------------|------------|-------|--------------------|------------|-----------|-----------|------------|------------|
| Ub | 7.03      | 0              | $\pi/2$    | $C_{lmax}$ | 40             | $100\pi$   | 0.7       | 1         | 1          | 1          |
| Lb | 5.743     | -12            | $-\pi/2$   | $-C_{lmax}$ | 5             | $1\pi$     | 0.1       | 0         | 0          | 0          |

*Table 7. Extended Solution domain for the aerocapture problem*

|    | $v(km/s)$ | $\beta\,(deg)$ | $\upsilon$ | $C_l$ | $\vartheta\,(deg)$ | $S\,(m^2)$ | $R_n/R_b$ | $m_{max}$ | $\sigma_1$ | $\sigma_2$ |
|----|-----------|----------------|------------|-------|--------------------|------------|-----------|-----------|------------|------------|
| Ub | 7.03      | 0              | $\pi/2$    | $C_{lmax}$ | 60             | $100\pi$   | 0.9       | 1         | 1          | 1          |
| Lb | 4.94      | -12            | $-\pi/2$   | $-C_{lmax}$ | 5             | $1\pi$     | 0.1       | 0         | 0          | 0          |

The Pareto front resulting from problem (41) and the restricted solution space, has been represented in Fig. 8, though there is some lack of resolution in some portions of the optimal set, in particular close to the zero of the design margins, the MACS approach has located a wide variety of solutions at different levels of belief. Notice how an increase of belief corresponds to an increase of propellant mass required to correct the manoeuvre. In the same way and increase of design margin causes an increase in the required mass margin.

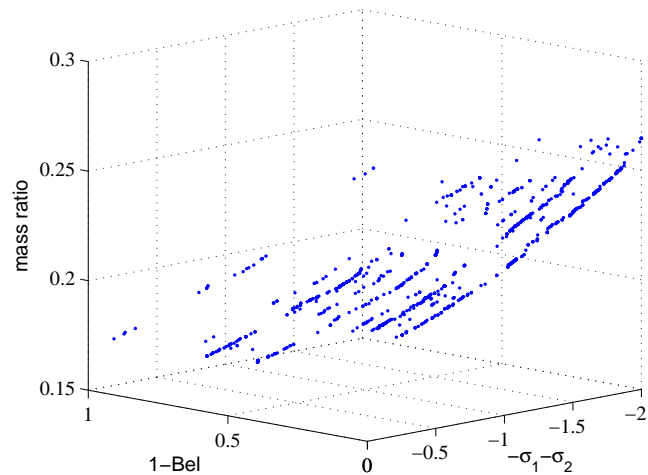

Figure 8. Three dimensional Pareto front of robust solutions

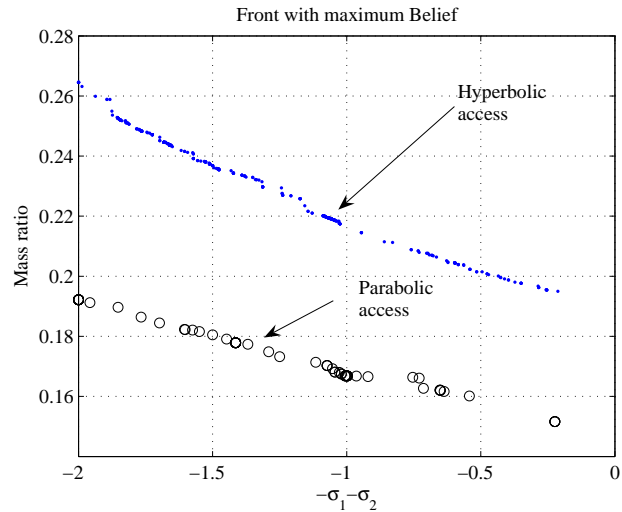

Figure 9. Pareto optimal fronts with maximum Belief for hyperbolic access (dotted line) and parabolic access to Mars (circle line)

A detail of the Pareto front with maximum belief is represented in Fig. 9 along with the analogous maximum belief front for the extended solutions pace. Notice how all the solutions of the latter front have a parabolic access and present a consistent improvement in the mass margin for all the values of the design margin.

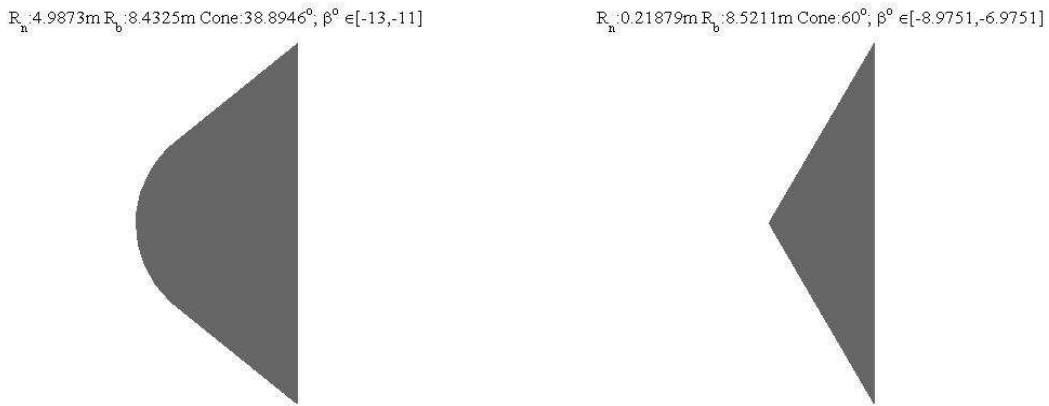

Figure 10. Examples of optimal shapes for two classes of arecapture manoeuvres: hyperbolic access on the left and parabolic access on the right

Figure 10 show two examples of solutions taken from the two fronts with maximum belief. In both cases the frontal area is considerably large but while former solution, left image of figure 10, the entry angle $\beta$ results to be quite steep, in the range [-13,-11] degrees, in the latter solution, right image of figure 10, the entry angle is quite shallow. This is consistent with the parabolic access velocity of the latter case.

All the solutions belonging to the maximum belief front fly at maximum lift in order to correct the error in the $\chi$ component. This is at the limit of validity of the aerodynamic model presented in this paper, therefore either a more sophisticated analysis would be required or a tight constraint on the maximum lift coefficient has to be inserted. This is the subject of an ongoing analysis.

## Final Remarks

In this paper the general reliability optimisation problem applied to a certain class of space mission design problems has been formulated using evidence theory for modelling the typical uncertainty existing during the early phases of the design process.

The resulting multiobjective optimisation process has been solved through a novel approach called multiagent collaborative search which blends together some of the main features of particle swarm optimization and classical evolutionary algorithms. In addition, an adaptive domain decomposition technique has been used to partition the solution domain and favour a more extensive exploration.

The application of the proposed methodology to a simple low-thrust test case has shown how robust solutions correctly fulfil all the constraints within the range of uncertainty of the delivered power. For a simple Earth to Mars transfer, a discontinuous set of robust solutions corresponding to different launch windows (short transfer, long transfer options) has shown a total mass of the combined propulsion and power subsystem of less than 0.266 of the total mass budget.

For the aerocapture test case, the first trivial result is that robust solutions exploit aerocapture as little as possible. The maximum belief front showed that solutions approaching the target planet with a low energy level result to be more robust than high energy ones. In both cases all solution resulted to be poorly sensitive to half cone angle, though configurations with a high reference area, high drag and low lift are more robust. This directly related to the absence of an active control during the atmospheric phase.

The design margin for a typical hyperbolic access to Mars taking into account the typical uncertainties on the atmosphere and on the aerodynamic properties resulted to be some 26% for a 2° entry corridor which is slightly over the typical 25% margin used in preliminary design for poorly known quantities.

# References

[1] Oberkampf W.L., Helton J.C. *Investigation of Evidence Theory for Engineering Applications*. AIAA 2002-1569, 4[th] Non-Deterministic Approaches Forum, 22-25 April 2002, Denver, Colorado

[2] Agarwal H., Renaud J., Preston E. *Trust Region Management Reliability Based Design Optimization Using Evidence Theory*. AIAA-2003-1779,5[th] AIAA Non-deterministic Approaches Forum, AIAA/ASME/ASCE/AHS/ASC Structures, Structural Dynamics and Materials Conference, Norfolk, Virginia, Apr. 7-10, 2003

[3] Vasile M. *A Systematic-Heuristic Approach for Space Trajectory Design*. Astrodynamics, Space Missions and Chaos, Ann NY Acad Sci 2004 Vol. 1017:234-254

[4] Vasile M., Summerer, L., De Pascale, P., *Design of Earth-Mars Transfer Trajectories using evolutionary branching technique*, Acta Astronautica 56 (2005) 705-720.

[5] De Pascale P., Vasile M., Finzi A.. *A Tool for Preliminary Design of Low-Thrust Gravity Assist Trajectories AAS/AIAA Spaceflight Mechanics Meeting,* AAS Paper 04-250, Maui Hi, 8-12 Feb.2004

[6] Schutze O., Mostaghim S., Dellnitz M., Teich J. *Covering Pareto Sets by Multilevel Evolutionary Subdivision Techniques*. Proceedings of EMO 2003, Faro, Portugal.

[7] Regan F.J., Anandakrishnan, *Dynamic of Atmospheric Re-Entry*, AIAA, Educational Series, 1993.

[8] Vinh N.X. Optimal Trajectories in Atmospheric Flight. Elsevier 1981.

[9] Deb K. Pratap A. Agarwal S. and Meyarivan. *A Fast Elitist Multi-Objective Genetic Algorithm: NSGA-II*. KanGAL Report No. 200001, 2000.

[10] Mitcheltree R.A., DiFulvio M., Horvath T.J.,Braun R.D. *Aerothermal Heating Predictions for Mars Microprobe*. Journal of Spacecraft and Rockets Vol.36,No.3,May-June 1999.

[11] Deb K. Pratap A. and Meyarivan T. *Constrained Test Problems for Multi-Objective Evolutionary Optimization.* KanGAL Report No. 200002, 2002.